\documentclass[12pt]{article}
\usepackage{times}
\usepackage{geometry}
\usepackage[square,comma,numbers,sort&compress]{natbib}
\usepackage[utf8]{inputenc}
\usepackage{enumitem,amssymb}
\newlist{thematic}{itemize}{8}
\setlist[thematic]{label=$\square$}
\usepackage{pifont}

\begin{document}

\bibliographystyle{unsrtnat}

\begin{flushleft}
\huge
Astro2020  Science White Paper \linebreak

The Discovery Potential of Space-Based Gravitational Wave Astronomy\linebreak
\normalsize

\noindent \textbf{Thematic Areas:} \hspace*{60pt} $\square$ Planetary Systems \hspace*{10pt} $\square$ Star and Planet Formation \hspace*{20pt}\linebreak
$\square$ Formation and Evolution of Compact Objects \hspace*{31pt} $\blacksquare$ Cosmology and Fundamental Physics \linebreak
  $\square$  Stars and Stellar Evolution \hspace*{1pt} $\square$ Resolved Stellar Populations and their Environments \hspace*{40pt} \linebreak
  $\square$    Galaxy Evolution   \hspace*{45pt} $\square$             Multi-Messenger Astronomy and Astrophysics \hspace*{65pt} \linebreak
  
\textbf{Principal Author:}

Name:	Neil Cornish
 \linebreak						
Institution:  eXtreme Gravity Institute, Department of Physics, Montana State University
 \linebreak
Email: ncornish@montana.edu
 \linebreak
Phone:  1 (406) 994 7986
 \linebreak
 
\textbf{Co-authors:} \\
Emanuele Berti, Physics \& Astronomy, Johns Hopkins University\\
Kelly Holley-Bockelmann, Astronomy, Vanderbuilt University \\
Shane Larson, Physics \& Astronomy,  Northwestern\\
Sean McWilliams, Physics \& Astronomy, West Virginia University\\
Guido Mueller, Physics, University of Florida\\
Priya Natarajan, Astronomy, Yale University\\
Michele Vallisneri, Physics, Caltech
  \linebreak

\textbf{Abstract :}

A space-based interferometer operating in the previously unexplored mHz gravitational band has tremendous discovery potential. If history is any guide, the opening of a new spectral band will lead to the discovery of entirely new sources and phenomena. The mHz band is ideally suited to exploring beyond standard model processes in the early universe, and with the sensitivities that can be reached with current technologies, the discovery space for exotic astrophysical systems is vast. 

\end{flushleft}

\pagebreak

\section*{Introduction:}

A gigameter scale space-based gravitational wave detector will open up four decades of the gravitational wave spectrum between 0.1 mHz and 1 Hz~\cite{Audley:2017drz}. Many sources are expected to populate this previously unexplored frequency band, including hundreds of black hole binaries across a wide range of masses and mass ratios, and thousands of frequency-resolved galactic binaries. If history is any guide, the opening of this new spectrum will also lead to the discovery of new astrophysical phenomena, just as the first X-ray satellite, Uhuru, provided the first compelling evidence for stellar mass black holes, and as the first gamma-ray satellites, Vela, discovered Gamma Ray Burst.

Gravitational waves are generated by the motion of all forms of mass/energy, and propagate unattenuated through the Universe from as far back as the Planck time. Energetic processes occuring in the very early Universe can generate high-frequency gravitational waves that will be redshifted into the mHz band today. A space-based gravitational wave detector will be sensitive to processes occuring in the first fractions of a second after the Big Bang, at energy scales between 1 TeV and 10 PeV, covering territory that is out of reach for current and proposed particle accelerators. These scales are of great interest for particle physics, covering phenomena such as the electroweak phase transition and the breaking of supersymmetry. A large number of potential cosmological sources of gravitational waves have been suggested, including cosmic strings~\cite{Siemens:2006yp}, novel inflationary scenarios~\cite{Bartolo:2016ami}, warped extra dimensions~\cite{Randall:2006py}, and various first-order phase transitions~\cite{PhysRevD.30.272,PhysRevLett.65.3080,Hogan:1986qda} that generate waves through bubble nucleation and collision, cavitation, percolation, and magnetohydrodynamic turbulence~\cite{PhysRevD.49.2837,Maggiore:1999vm}.
		
Hitherto unknown astrophysical phenomena may also generate detectable levels of gravitational waves. Possibilities include binary systems of Boson stars~\cite{Palenzuela:2007dm,Eby:2015hsq}, the inspiral of stellar-mass compact objects into supermassive Boson stars~\cite{Kesden:2004qx}, the motion of clumps of self-interacting dark matter~\cite{Pollack:2014rja, Hall:2016usm}, the collapse of supermassive stars~\cite{Shibata:2016vzw}, and the super-radiant excitation of bosonic condensates~\cite{Brito:2017zvb}. Very likely, the exotic sources detected by a space-based observatory have yet to be imagined. 

Detecting unexpected signals is challenging since we can not build templates to extract the signals from the instrument noise. And with a single triangular laser interferometer we can not cross correlate the signals from multiple independent detectors as is done with ground-based interferometers and pulsar timing arrays. However, methods have been developed that will allow us to detect unknown signals with a single space based interferometer: For signals with periods longer than the light travel time between the spacecraft, the laser phase measurements can be combined to form a signal-insensitive data combination that can be used to infer the noise in the signal-sensitive data channels and detect excess signal power~\cite{Tinto:2001ii}; while for short period signals the arrival times at each spacecraft can be used to separate signals from noise~\cite{Robson:2018jly}.

\section*{Cosmological Sources:}

One of the most intriguing fundamental discoveries in the mHz band would be evidence that the physical vacuum could have undergone one or more phase transitions in the very early universe. Symmetry breaking transitions that lead to first-order phase transitions nucleate bubbles of the new vacuum, which then expand and collide, creating plasma turbulence and sound waves. These coherent plasma motions can be a potent source of gravitational radiation, driving a substantial fraction of mass in the universe at relativistic speeds. Many extensions of the standard model predict a first-order phase transition associated with electroweak symmetry breaking at the TeV scale~\cite{PhysRevD.30.272,PhysRevLett.65.3080,Hogan:1986qda,PhysRevD.94.075008,ESPINOSA2012592,Kozaczuk:2014kva} that would produce gravitational waves in the mHz band today. Supersymmetric extensions predict additional phase transitions at higher energies~\cite{Kozaczuk:2014kva} that may also be detectable by a space-based gravitational wave detector. The frequencies of the gravitational waves produced in these transitions is set by the Hubble parameter at the time of the transition (which has units of Hz). These high-frequency gravitational waves are subsequently redshifted into the mHz band by the expansion of the universe.

While the standard model of particle physics does not predict a first-order electroweak phase transition~\cite{PhysRevLett.77.2887,PhysRevLett.113.082001}, we know that the theory is incomplete, and there are good reasons to think that the transition is indeed first order, including the need to break the matter - anti-matter symmetry, the need for an out-of-equilibrium environment necessary for baryogenesis to occur~\cite{annurev.ns.43.120193.000331,Katz:2016adq}, and a mechanism to generate seed magnetic fields~\cite{Durrer2013,Subramanian_2016,PhysRevD.96.123528}. The discovery of gravitational waves from the electroweak phase transition would be monumental, with profound implications for fundamental physics and the origin of all matter in the universe.
			
In addition to phase transitions, many other exotic sources of gravitational waves may produce detectable signals in the mHz band, including cosmic superstrings, macroscopic extra dimensions, and TeV-PeV scale reheating after inflation.

\section*{Astrophysical Sources:}

To get a sense of the types of exotic objects that could be detected by a space-based gravitational wave detector, it is useful to consider the strength of the signal as a function of the energetics of the source. A gravitational wave signal detectable at cosmological distances requires large amounts of mass moving at close to the speed of light. A good example of this is a merging supermassive black hole binary, where millions of solar masses collide at half the speed of light. On the other hand, nearby sources can be detected even if they are slowly moving and not very massive. A good example of this is a galactic white dwarf binary, where one solar mass moves at one-thousandth the speed of light.
 
Thus, nearby sources do not have to be very massive or relativistic to be detected by a space-based gravitational wave observatory, though they will have to be fairly common to yield a reasonable number of detections within the small cosmological volume in which they can be detected. Conversely, highly relativistic and massive sources can be detected throughout the Universe, so there is a good potential for discovery, even if the events are rare.


A number of exotic sources of mHz-band gravitational waves have been proposed. One possibility is binary systems of horizonless objects such as boson stars~\cite{Palenzuela:2007dm,Eby:2015hsq} or ``fuzz balls''~\cite{Mathur:2005zp}. These could potentially be distinguished from binary black hole mergers due to the excitation of interior oscillation modes, and from the modified dynamics of the merger and ringdown. Other possibilities include hyperbolic encounters between compact objects~\cite{Capozziello:2008mn, DeVittori:2012re}, and the collapse of supermassive stars~\cite{Shibata:2016vzw}. 
Another intriguing possibility is to turn spinning black holes into a dark matter detector that is sensitive to a particular class of dark matter candidates. The idea is that ultralight bosons can tap the rotational energy of spinning black holes via a superradiant instability that generates a nearly monochromatic stochastic gravitational wave background that peaks at a frequency set by the mass of the boson~\cite{Brito:2017zvb}.

\section*{Detection:}

Signal templates - detailed models for the gravitational waveforms found by solving Einstein's equations for the source - provide the most powerful method for distinguishing signals from noise and inferring the physical properties of the sources. But all is not lost when templates are not available. Indeed, all of the signals detected by LIGO through the first and second observing runs, save for GW151226, were picked up by minimally-modelled `burst' searches that reconstruct the signals using wavelet representations~\cite{Abbott:2016ezn,TheLIGOScientific:2016uux}. In addition to detecting signals, these burst methods are able to infer the signal duration, frequency content, and luminosity evolution of the source~\cite{Cornish:2014kda}. With a network of detectors, cross-correlation techniques, which exploit the fact that the signals will be correlated across the network while the noise will (hopefully) not be correlated, can be used to detect a wide range of signal types, including stochastic signals~\cite{Romano:2016dpx}.

Detecting un-modeled sources with a single space-based detector will be more challenging than with a network of interferometers on the Earth, or with an array of pulsars. The three arms of the detector can be used to synthesize three (almost) independent data channels, two of which measure the plus and cross gravitational wave polarization states, and a third which is insensitive to gravitational waves with periods less than the light travel time between the spacecraft~\cite{Tinto:2001ii}. These three data channels are signal-orthogonal, so cross correlation between channels can not be used for detection, and in any case, the channels will have correlated noise. The signal-insensitive channel can be used to estimate the instrument noise level in the signal-sensitive channels~\cite{Hogan:2001jn,Adams:2010vc,Adams:2013qma}, and to veto noise transients, making it possible to detect excess power from stochastic gravitational waves, and to distinguish gravitational wave bursts from noise disturbances~\cite{Robson:2018jly}. At higher frequencies, where all three channels are equally sensitive to gravitational waves, the difference in the temporal signature of gravitational waves and noise disturbances in the data allows the effects to be separated~\cite{Robson:2018jly} (noise always appears with delays equal to the light travel time between spacecraft, while signals have delays that depend on the sky location of the source). 

In both instances, it is the discriminatory power provided by the three-arm design that makes it possible to detect un-modelled and exotic sources, and to open up the full discovery potential of space-based gravitational wave astronomy. 

\section*{Summary:}

A space-based interferometer operating in the previously unexplored mHz gravitational band has tremendous discovery potential. If history is any guide, the opening of a new spectral band will lead to the discovery of entirely new sources and phenomena. The mHz band is ideally suited to exploring beyond standard model processes in the early universe, and with the sensitivities that can be reached with current technologies, the discovery space for exotic astrophysical systems is vast. 

\newpage

\bibliography{LISA}

\end{document}